# The role of micro-credentials in the future digitalized AI-driven education

**Wadim Strielkowski**
Prague Business School
Prague, Czech Republic
e-mail: strielkowski@pbs-education.cz

**Abstract**

This paper focuses on assessing the potential of micro-credentials in digitalized higher education in the era of artificial intelligence (AI). Micro-credentials are commonly described as "mini qualifications" or "digital badges" that certify an individual's competency in a specific skill or cluster of skills. Being different from traditional academic credentials in both their granularity and focus, micro-credentials represent a new and exciting topic that is rapidly gaining popularity all around the world both within higher educational institutions (HEIs), such as public and private universities, as well as within other education providers, such as non-governmental organizations and SMEs. The popularity of micro-credentials has been further enhanced by the recent COVID-19 pandemic which caused digital surge in higher education leading to many rapid technological innovations and changes that were unthinkable before. The rising interest in micro-credentialing can be best demonstrated by the increase of the number of scientific publications on this topic from just 1 in 1992 to 165 in 2024.
The paper employs a comprehensive bibliometric network analysis using the terms "micro-credentials" based on a sample of 608 selected publications indexed in the Scopus database. It carries out the network cluster analysis using both text data as well as bibliometric data with the help of VOSviewer software. The results and outcomes of this research might be helpful for researchers, stakeholders, and policymakers in devising effective strategies and policies for transforming the future digitalized AI-driven higher education.

**Keywords:** Micro-credentials, higher education, digital transformation, AI, bibliographic analysis

## 1. Introduction

Micro-credentials are emerging as a significant innovation in education, representing focused certifications of specific competencies often conferred via digital badges or similar formats [1,2]. For example, UNESCO describes a micro-credential as a record of focused learning achievement that verifies what learners know or can do, typically awarded as a badge or certificate by a trusted provider and meeting clear standards [3]. Micro-credentials are smaller in scope than traditional degrees and diplomas - being more granular they concentrate on specific skills or even clusters of skills. This granularity makes them especially adaptable to the rapidly changing skill demands of today's economy [4]. In the era of digital transformation and artificial intelligence (AI), education systems are under pressure to become more flexible, personalized, and competency based [5-8]. Micro-credentials represent a good way to certify skills in a timely, modular fashion, aligning with the needs for the AI-driven lifelong learning.

In recent years, several trends contributed to the rising popularity of micro-credentials. The COVID-19 pandemic accelerated the digitalization of higher education, forcing higher education institutions (HEIs), universities, and learners to adopt online platforms and new modes of learning virtually overnight [9-11]. This digital surge in academia and higher education not only familiarized millions with online learning but also highlighted the shortcomings of traditional credentialing in recognizing skills obtained outside formal degree programs. Even before the pandemic employers worldwide were making concerns that graduates lacked practical skills and job-ready competencies, despite holding traditional degrees [12, 13]. High tuition costs and student debt have raised questions about the exclusive value of traditional degree programs, while many adult learners and working professionals seek shorter, just-in-time learning options that can boost their employability. As a result, governments, universities, and the corporate sector began to rethink the traditional credentials schemes moving beyond the monopoly of the traditional university degrees toward more flexible credentials such as digital certificates, badges, as well as micro-credentials [14, 15].

Therefore, the intersection of digital transformation, AI, and evolving workforce needs has created suitable grounds for micro-credentials in higher education and beyond. This paper identifies key trends, common challenges, and opportunities in the role of micro-credentials in the future digitalized and AI-driven education through bibliometric analysis using a sample of publications indexed in Elsevier's Scopus database. It analyzes publication trends employing a bibliographic network analysis approach and discusses what its results reveal about the future of AI-driven education.

## 2. Micro-credentials in higher education and beyond

In the last few years, academic interest in micro-credentials has grown exponentially reflecting the rapid evolution of this concept to a global education trend. Early literature often focused on digital badges which represented a form of micro-credential pioneered in the 2010s to certify skills and learning processes outside the formal curriculum [16, 17]. Some researchers note that while employers value the granular skill information that badges provide, questions arose about their validity compared to degrees [18]. Over time, however, studies began to affirm the potential of micro-credentials to demonstrate what a candidate knows and can do in a concrete way that traditional transcripts do not [19].

Most recently, research literature focused on studying micro-credentials as tools for competency-based education, lifelong learning, and curriculum innovation. Some studies demonstrate that integrating micro-credentials into higher education curricula can increase flexibility and help direct programs toward competence-based outcomes aligned with job requirements [20]. This means that universities experimenting with embedding micro-credential modules within degrees or offering stand-alone certifications in high-demand skills (e.g. data analytics, project management) that can later stack into larger qualifications [21]. The research suggests that such approaches can better motivate students, allowing them to pursue topics of personal and professional interest while also enhancing their employability with tangible skill endorsements.

Another dominant theme in research literature is the mismatch between traditional education and labor market needs, and how micro-credentials might address it. Many scholars have documented employers' complaints that graduates often lack practical and technical skills, even as jobs are rapidly changing due to automation and AI [22-24]. There is no secret that many HEIs and universities nowadays still tend to adapt the old-fashioned ("before the Internet") approach of making the students memorize information and then examine whether it has been memorized in a satisfactory way. Formal degrees are sometimes criticized for being slow to adapt to their curricula, leaving a gap between what is taught and the competencies in demand. By contrast, micro-credentials can be developed quickly to meet emerging skill requirements [25, 26]. For example, they can provide certifications in new programming languages, AI tools, or data literacy and thus act as viable supplements to a university degree.

Figure 1 that follows shows the dynamics of frequency of search (called "Interest over Time") of the keywords "micro credential", "micro degree", and "digital badge" using Google Trends research tool and covering a time span from 2015 until 2025. One can clearly see a visible surge in these terms during the early stages of the COVID-19 pandemic as well as most recently (second half of 2025) which only confirms the growing interest in micro-credentialing.

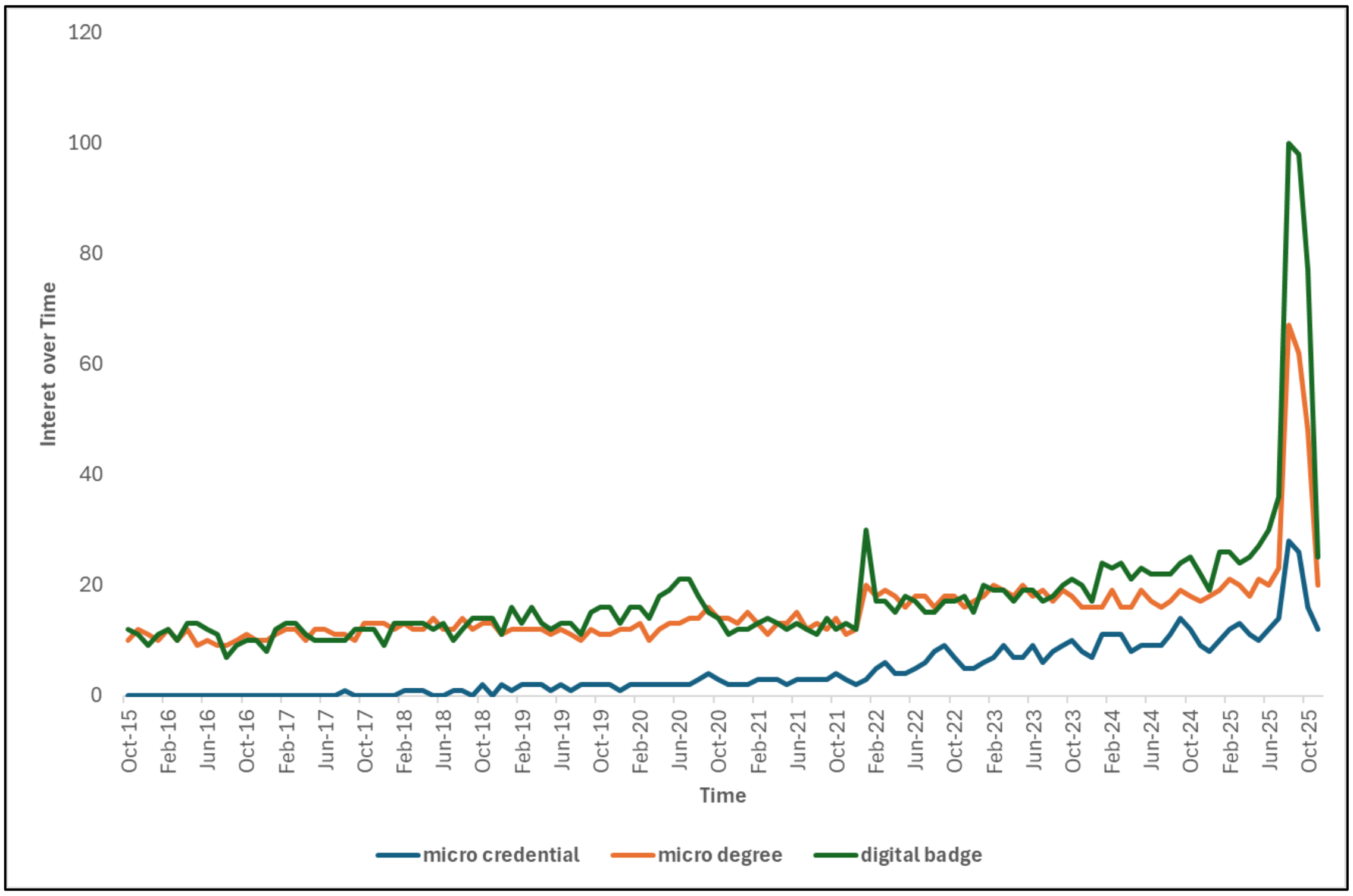

**Fig. 1.** Dynamics of frequency of search of the terms of "micro credential", "micro degree", and "digital badge" (2015-2025)
Source: Own results based on Google Trends

Furthermore, quality assurance and frameworks constitute a major concern for micro-credentials. Because dozens of definitions and practices have emerged across different countries and institutions, a challenge noted in the literature is establishing common standards for what a micro-credential represents [27]. Many researchers explore the use of Blockchain technology to secure and verify micro-credentials, ensuring they are verifiable and easily shareable across platforms [28].

Finally, many studies focus on the fact that micro-credentials are not only the domain of universities. Non-HEIs and non-university actors such as industry groups, startups, non-profits, and small and medium enterprises (SMEs) have become active in credentialing [29, 30]. This represents a completely new approach that now challenges the historical entrenched monopoly of higher education institutions in certifying learning.

## 3. Data

This study is based on a bibliometric network analysis of publications related to micro-credentials indexed in Elsevier's Scopus database. Our dataset consists of 608 publications indexed between 1992 and 2025, all of which contain the keyword "micro-credentials" (or a closely related term) in their title, abstract, or keywords. The decision to use Scopus as the main source of data, rather than the Web of Science (WoS) or other databases, was motivated by Scopus's broader coverage in the social sciences and education literature. Scopus indexes a larger number of journals and includes more conference proceedings in these fields, which makes it a more comprehensive repository for an emerging, interdisciplinary topic such as micro-credentials. Figure 2 that follows shows the trend of the publications on micro-credentialing over time (2002-2024).

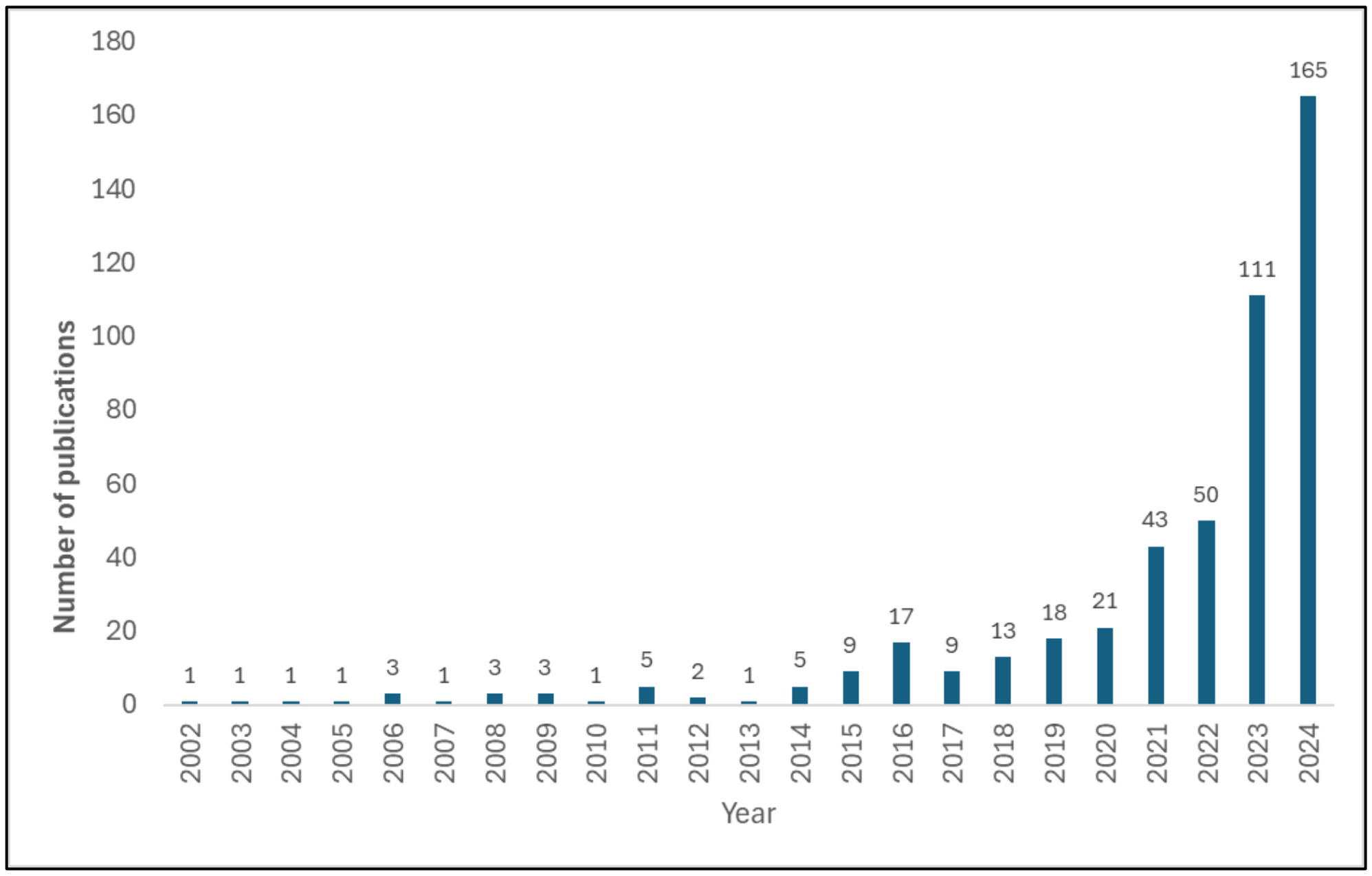

**Fig. 2.** Trend of the publications on micro-credentialing over time (2002-2025)
Source: Own results based on Scopus database

Figure 3 depicts a Word Cloud diagram created using the titles of 608 publications related to micro-credentialing retrieved from Scopus database and highlighting the occurrence and frequency of keywords connected to various types of research on micro-credentials. The larger and bold keywords demonstrate higher usage frequency.

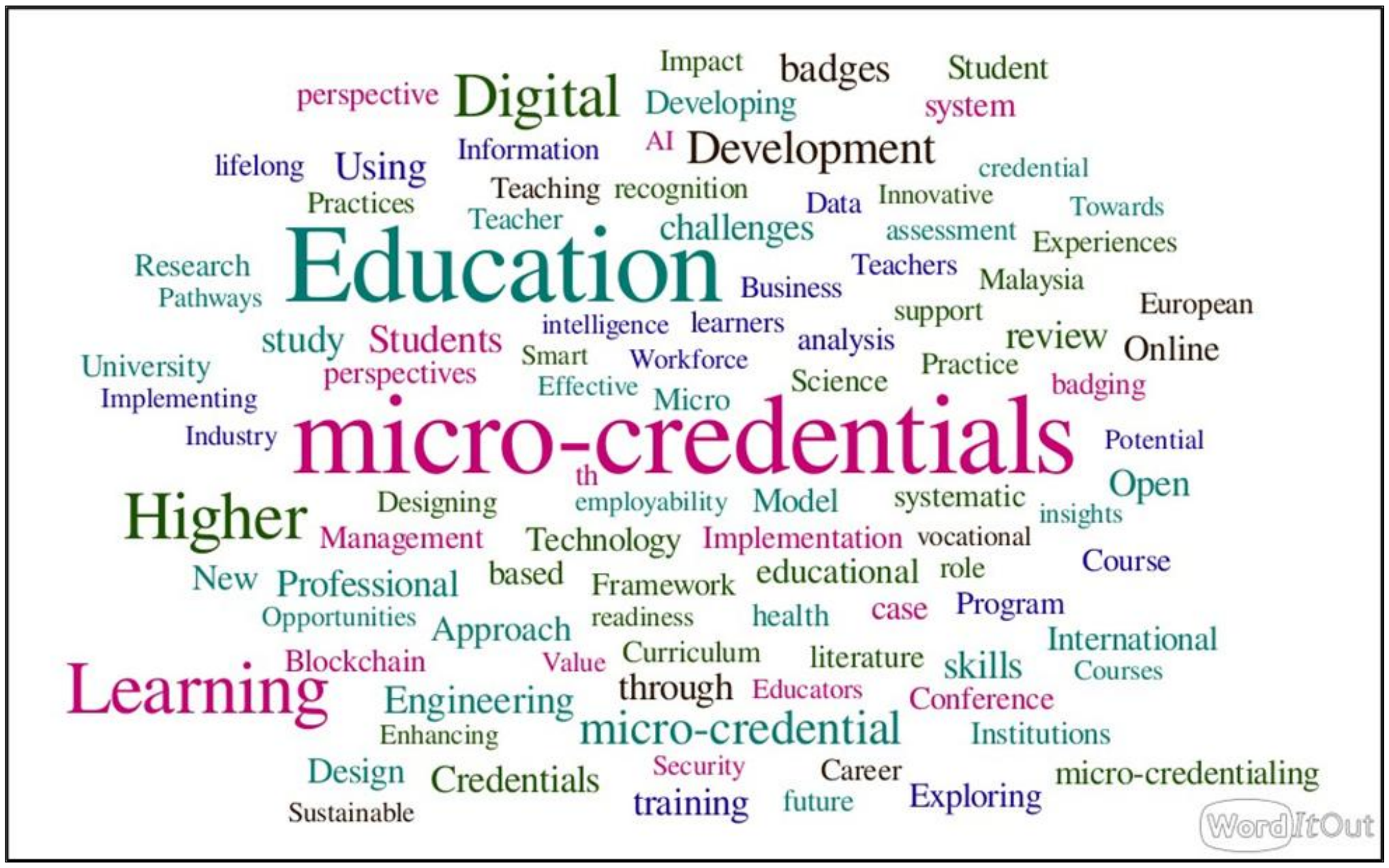

**Fig. 3.** Word Cloud based on the titles of 608 papers from Scopus database (1992-2025)
Source: Own results using WordItOut

## 4. Materials and methods

A bibliometric network analysis was employed to examine the intellectual structure of micro-credential research. This allowed to quantify publication patterns and visualize relationships (such as co-occurrence of keywords or co-citation of articles) within the vast body of literature. VOSviewer software (version 1.6.18) was employed to conduct this analysis, which is a widely used tool for constructing and visualizing bibliometric networks. VOSviewer is well-suited for revealing clusters of related items (e.g. groups of papers, authors, or terms that frequently appear together), helping us identify the main thematic areas ("clusters") in the micro-credentials discourse.

Table 1 that follows shows the summary of data and the selection algorithm reporting on the document types while Figure 4 offers an outline of the applied bibliometric research methodology.

**Table 1.** Summary of data and data selection algorithm

| Category | Specific criteria |
|---|---|
| Reference and citation database | Scopus |
| Time period | 1992-2025 |
| Language | "English" |
| Keywords | "micro-credentials" |
| Document types: | |
| Articles | 272 |
| Proceeding papers | 183 |
| Others | 153 |
| Sample size | N = 608 |

Source: Own results

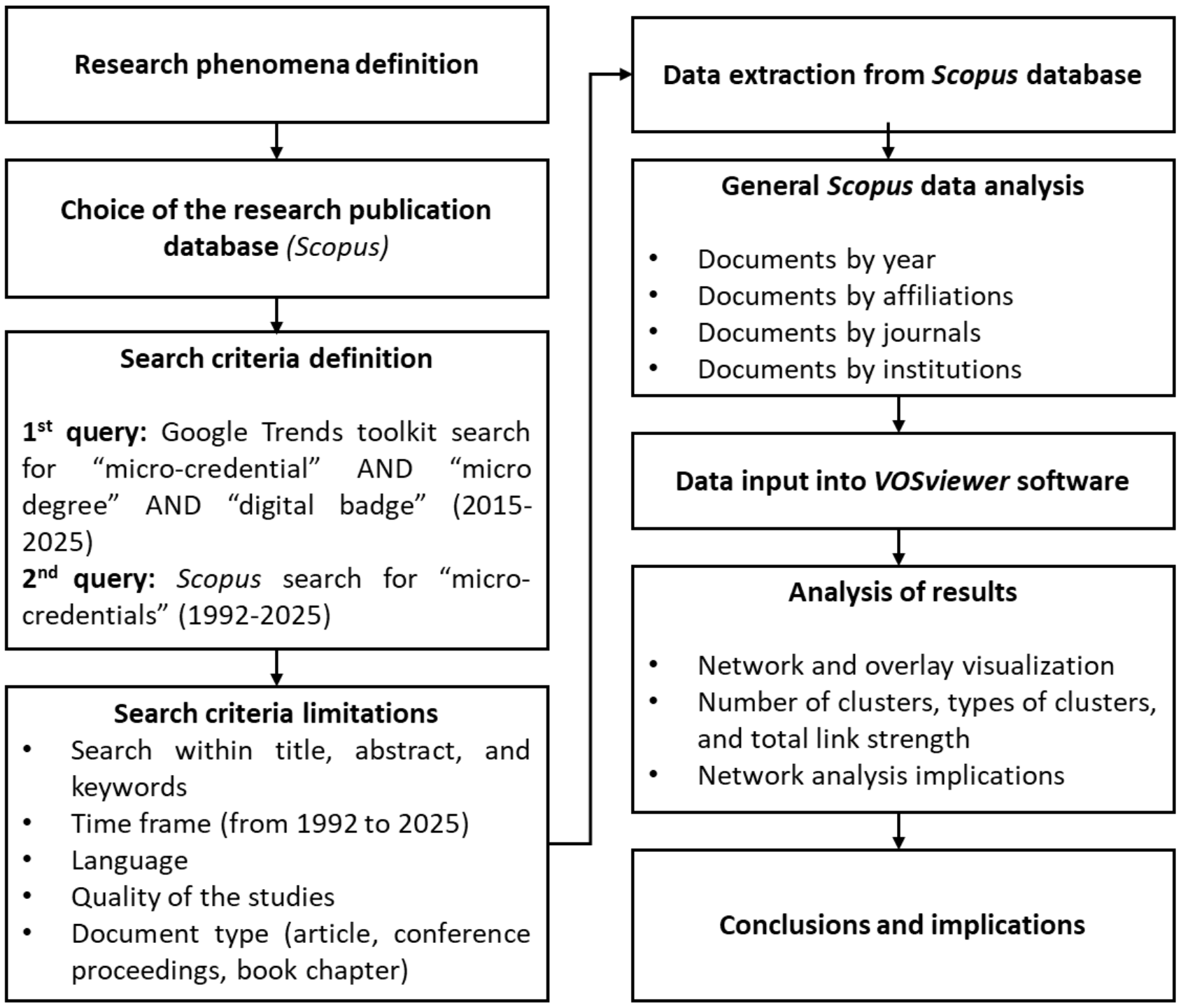

**Fig. 4.** Outline of the bibliometric research methodology.
Source: Own results

## 5. Results and discussion

In this section, we present the results of the empirical model based on the bibliometric analysis that employs the VOSviewer software tool. The results of the analysis are presented in the form of visual network maps that allow for deducing the key patterns and occurrences.

Figure 5 below presents the visualization of the network cluster analysis with a map based on the text data from the sample of all 608 publications indexed in Scopus database from 1992 until 2025. Our results of the bibliometric network analysis demonstrate that five main clusters were identified.

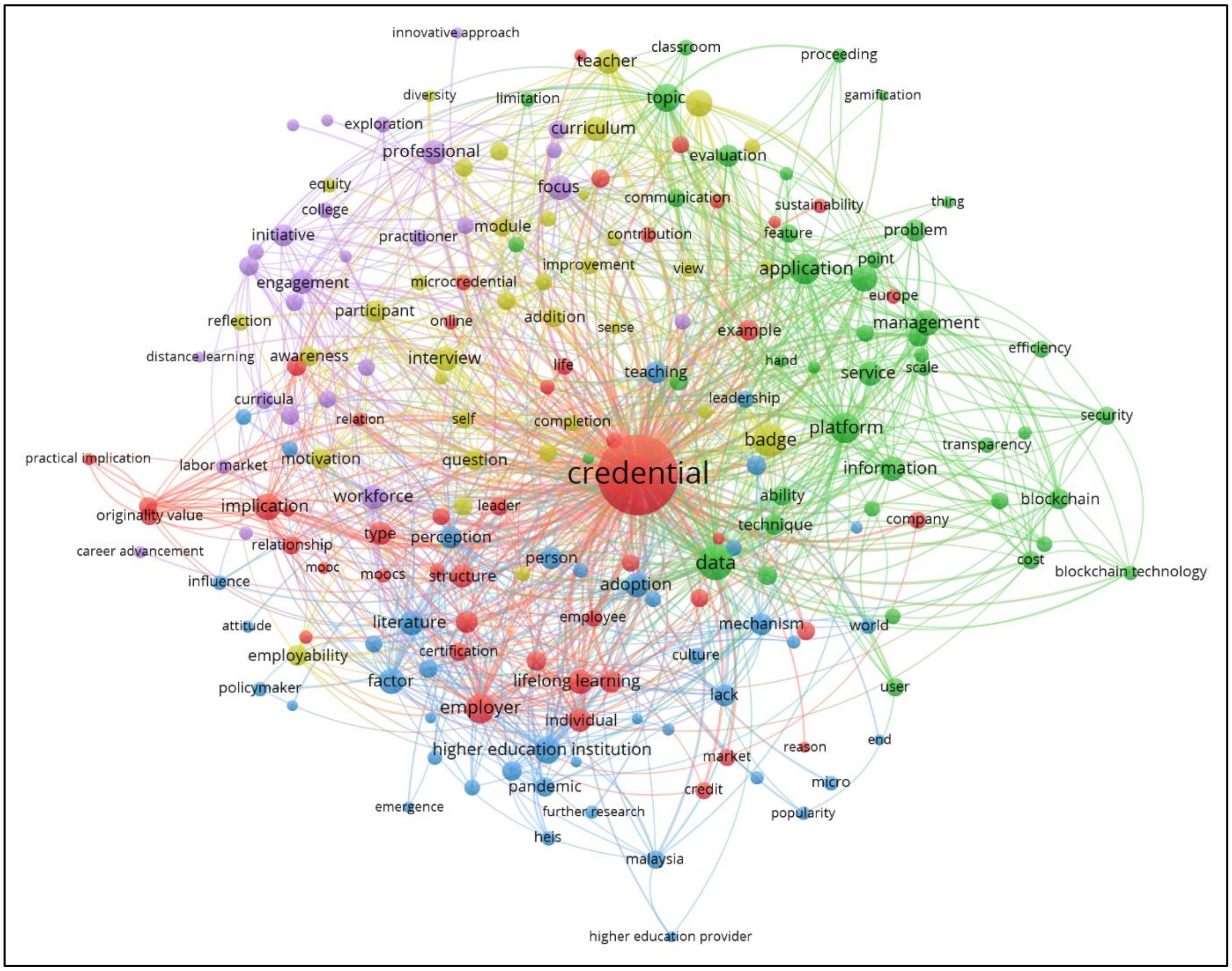

**Figure 5.** The dominant clusters of cross-sector research connected with micro-credentials retrieved from the sample of 608 publications indexed in Scopus.
Source: Own results based on VOSViewer v. 1.6.18 software

The analysis of using keywords and phrases in the publications revealed that the key terms connected to "micro-credentials" are most often associated with the following concepts: i) employability and micro-credentials; ii) distance learning and flexible education; iii) learning and skills development; iv) digitalization and technology integration; v) quality assurance and sustainability.

Our results demonstrate that lifelong learning emerges here as a cross-cutting concept linking multiple clusters. It appears in the employability cluster (as workers continuously update skills) and in the distance learning cluster (as flexible online courses enable ongoing education). This indicates that micro-credentials serve both immediate labor market needs and long-term personal development goals. Similarly, digital badges are discussed in both the distance learning cluster and the technology cluster, underscoring how technological tools (such as digital badges or Blockchain) highlight the success of online micro-credential programs. The intersections between clusters suggest that effective micro-credential initiatives often integrate multiple dimensions: for instance, a program might leverage technology (Cluster 4) to deliver online courses (Cluster 2) that enhance learners' skills and motivation (Cluster 3) for better employment outcomes (Cluster 1) while adhering to quality standards (Cluster 5). Thence, higher education and employment appear to be strongly entangled in the micro-credentialing debate. Indeed, the keyword analysis confirms that micro-credentials have strong links with higher education, digital badges, employment and lifelong learning. This reflects a consensus in the literature that micro-credentials act as a bridge between formal education and the skills demanded by the workforce, all facilitated by digital technologies and guided by quality considerations.

Furthermore, Figure 6 reveals the results of the network map based on the bibliographic data (keyword co-occurrences, citation, and bibliographic coupling) retrieved from 608 publications selected from Scopus database (1992-2025). Our results of the bibliometric network analysis demonstrate that six main clusters were identified.

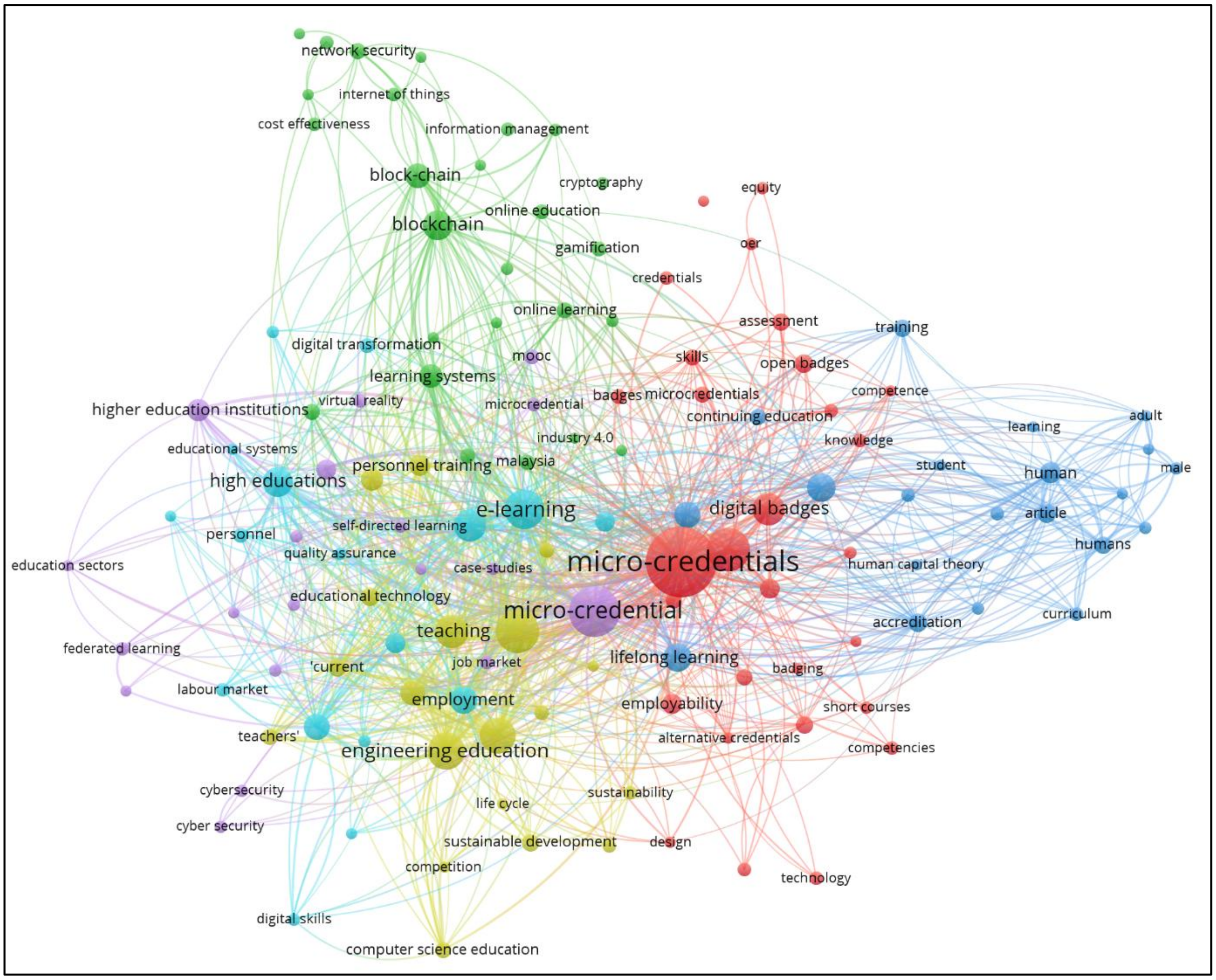

**Figure 6.** Network map based on the bibliographic data of the sample of papers containing the keywords "micro-credentials" retrieved from the sample of 608 publications indexed in WoS.
Source: Own results based on VOSViewer v. 1.6.18 software.

The clusters in question are the following: i) learner motivation and benefits; ii) alternative credentialing (digital badges & MOOCs); iii) lifelong learning and continuing education; iv) AI-enhanced learning; v) teacher professional development; and vi) equity and inclusion in education.

The six clusters from Figure 6 demonstrate the broad scope and interdisciplinary nature of micro-credential research. In comparison to the text-based clusters, the bibliographic clusters place additional emphasis on emerging technology (Cluster 4 on AI) and specific use-cases like teacher training (Cluster 5), while still covering core themes of motivation, alternative credentials, lifelong learning, and inclusion. Notably, these themes are interconnected: for instance, integrating AI in micro-credential platforms (Cluster 4) could enhance personalized learning, which in turn might improve learner motivation and outcomes (related to Cluster 1). Similarly, efforts to use micro-credentials for teacher development (Cluster 5) have implications for scaling quality education and inclusion (Cluster 6), since well-trained teachers are crucial for educational equity. Thus, our bibliographic analysis reinforces and extends the insights from Figure 5 demonstrating that micro-credentials are a multifaceted phenomenon, being studied as technological innovations, pedagogical tools, workforce development mechanisms, and policy-relevant interventions all at once. The identification of six distinct yet related clusters in Figure 6 confirms that the field of micro-credentials in higher education is rapidly evolving and expansive, touching on individual, institutional, and societal levels of impact.

## 6. Conclusions and implications

Overall, this study clearly shows that micro-credentials represent transformative tools at the crossroads of education, technology, and the labor market. Our bibliometric analysis confirmed the multifaceted role that micro-credentials play in the future digitalized AI-driven education ecosystem. Key concepts such as higher education, digital badges, employment, and lifelong learning emerge as common threads, indicating that micro-credentials form a nexus between formal educational programs and ongoing professional skill development. In practical terms, this means micro-credentials are increasingly seen to make education more flexible and responsive to individual and industry needs. They support learners in obtaining targeted skills and on-demand certifications, while also enabling employers to recognize those skills in a transparent, digital form.

Several important implications can be drawn from our bibliometric network cluster analysis. The prominence of the employability theme and the learner motivation theme highlights that micro-credentials are valued for improving job prospects and aligning education with market needs. Policymakers and higher education institutions should recognize micro-credentials as a strategy to bridge the gap between academic curricula and workforce competencies. This may involve closer collaboration with industry to ensure that the content of micro-credential programs directly addresses current skill gaps and competency development" priorities. When implemented well, micro-credentials can enhance human capital by enabling continuous upskilling, thereby making the labor force more adaptable in the face of rapid technological change.

In addition, this paper confirms that digital technologies (including online learning platforms, Blockchain, and AI) are integral to the micro-credential landscape. Institutions should invest in robust digital infrastructure for issuing and managing micro-credentials, such as Blockchain-based credentialing systems that ensure security and global verifiability of qualifications. Moreover, the emergence of an AI-focused cluster suggests that AI can be used to personalize learning pathways and provide intelligent support in micro-credential courses. For example, AI-driven adaptive learning systems could tailor micro-credential content to individual learner needs, thereby improving engagement and completion rates. Embracing these technologies will be crucial for scaling up micro-credential offerings in a cost-effective and learner-centric way as higher education becomes more digitalized and data-driven.

Furthermore, it appears that the long-term success of micro-credentials depends on trust and recognition across educational and professional spheres. Stakeholders must develop clear standards and frameworks for micro-credentials – for instance, defining learning outcomes, assessment rigor, and credit transfer mechanisms, so that these credentials are widely accepted. Quality assurance measures (such as accreditation or endorsement by reputable institutions) will increase the credibility of micro-credentials in the eyes of employers and learners. Additionally, aligning micro-credential programs with broader initiatives such as the Sustainable Development Goals (SDGs) can ensure they contribute to long-term well-being of individuals and societies, rather than isolated training episodes. In practice, this means designing micro-credentials that not only teach specific skills but also uphold values of equity and cultural inclusivity, thereby supporting diverse learners. Ensuring equal access is part of quality - digital credentials should be accessible to learners from various backgrounds, including those in underrepresented or disadvantaged groups, so that micro-credentials serve as a tool for social inclusion and not a source of new disparities.

In addition, micro-credentials can transform not only student learning but also faculty development and adult education practices. Educational authorities and universities might consider integrating micro-credential pathways for teachers and academic staff to continuously update their pedagogical skills (for example, in using educational technology or new teaching methods). This creates a ripple effect – educators who earn micro-credentials become ambassadors of this model, incorporating micro-credentialing in their own teaching and thereby normalizing it for students. Furthermore, governments and organizations should embed micro-credentials into lifelong learning frameworks (e.g., professional licensing, continuing education requirements) to encourage a culture where individuals regularly return to learning new skills. By doing so, micro-credentials can become a cornerstone of lifelong learning systems, enabling workers to remain employable and societies to stay competitive amid evolving skill demands.

Finally, our findings suggest that research and implementation of micro-credentials have been concentrated in certain regions (North America, Europe, and Australia) and within specific communities of practice. As interest in micro-credentials grows worldwide, there is a need for knowledge-sharing and collaboration across countries and sectors. International standards bodies and cross-sector partnerships (involving universities, industry certification bodies, and government agencies) could play a role in harmonizing micro-credential frameworks, so that a credential earned in one context is understood and valued in another. Our analysis points to common goals across the globe – improving education-employment linkages, leveraging technology in learning, and expanding access which can serve as a foundation for international dialogue. By working together, stakeholders can devise effective strategies and policies to fully realize the potential of micro-credentials in transforming higher education. Such strategies may include creating interoperable platforms for credential exchange, adopting open badge standards, or developing guidelines for accrediting micro-credential programs across institutions and borders.

Micro-credentials have clearly emerged as a promising innovation to address the challenges and opportunities of an AI-driven, digital higher education environment. They represent a shift toward more modular, skills-focused learning that complement traditional degrees. However, realizing their promise requires careful attention to motivation and engagement of learners, technological infrastructure (including AI integration), quality assurance and recognition, and alignment with societal goals such as equity and lifelong learning. Researchers, educators, and policymakers alike need to devise informed strategies for integrating micro-credentials into the future of education. If properly supported by robust policy and collaboration, micro-credentials can become a key component in transforming higher education to be more flexible, inclusive, and responsive to the needs of learners in the 21st century.